\documentclass[a4paper,11pt]{article}
\usepackage{makecell}
\usepackage{pos}
\usepackage{lipsum} 
\usepackage{subfigure}
\title{Summary of IceCube Tau Neutrino Searches and Flavor Composition Measurements of the Diffuse Astrophysical Neutrino Flux}

\ShortTitle{Summary of Tau Neutrino Searches in IceCube}

\author{The IceCube Collaboration \\{\normalsize \normalfont(a complete list of authors can be found at the end of the proceedings)}\\}

\emailAdd{neha.lad@desy.de}
\emailAdd{cowen@phys.psu.edu}

\abstract{

We present a summary of the flavor composition measurements for the diffuse astrophysical neutrino flux using data from the IceCube Neutrino Observatory at the South Pole. IceCube has identified candidate astrophysical tau neutrinos through two different approaches. One approach used a dedicated particle identification algorithm for the classification and reconstruction of the 'Double Cascade' event topology, a signature of tau neutrino charged current interactions. This first approach is applied to the High Energy Starting Events (HESE) sample, an all-sky, all-flavor set of neutrino events with energy above 60~TeV encompassing 12 years of IceCube livetime. We show that the addition of more years of data and updated ice properties on the HESE sample delivers tighter constraints on the flavor composition of the astrophysical neutrino flux than previous IceCube analyses, in particular when it is fit in combination with high statistics samples of through-going tracks and cascades.  A second approach uses a sensitive machine-learning-based selection technique that finds seven candidate events in 9.7 years of IceCube data.  This approach excludes the zero astrophysical tau neutrino hypothesis at the highest statistical significance to date.

\vspace{4mm}
{\bfseries Corresponding authors:}
Neha Lad$^{1,2,*}$, D.~F.~Cowen$^{3,4}$\\
{$^{1}$ \itshape Deutsches Elektronen-Synchrotron DESY: Zeuthen, Germany}\\
{$^{2}$ \itshape Institut für Physik, Humboldt-Universität zu Berlin: Berlin, Germany}\\
{$^{3}$ \itshape Dept. of Physics, Pennsylvania State University, University Park, PA, USA}\\
{$^{4}$ \itshape Dept. of Astronomy and Astrophysics, Pennsylvania State University, University Park, PA, USA}\\
[4mm]
$^*$ Presenter

\ConferenceLogo{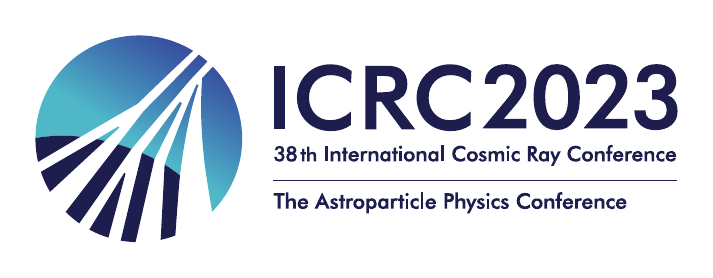}

\FullConference{The 38th International Cosmic Ray Conference (ICRC2023)\\ 26 July -- 3 August, 2023\\ Nagoya, Japan}
}

\begin{document}

\maketitle

\section{Introduction}\label{sec1}
The IceCube Neutrino Observatory~\cite{Aartsen:2016nxy}, buried deep in the ice at the South Pole, has played a pivotal role in the study of the astrophysical diffuse neutrino flux by reporting evidence of diffuse astrophysical neutrinos~\cite{discovery_paper}. However, this measurement alone does not give a complete picture of the astrophysical sources or the mechanisms by which they generate neutrinos. A study measuring the neutrino flavor content can thus enhance our understanding of various high-energy particle factories of the universe. 

By employing an array of 5,160 digital optical modules (DOMs) buried in the ice, IceCube detects the flashes of light produced by neutrino interactions. These interactions can create different signatures in IceCube based on the type and flavor involved. Single cascades can be created by charged current (CC) electron neutrino ($\nu_{e}$) and neutral current (NC) interactions of any flavor. Light depositions along a long track traversing the detector are created by muon neutrino ($\nu_\mu$) CC interactions and atmospheric muons. Tau neutrino ($\nu_{\tau}$) interactions can be distinguished from other neutrino interactions, provided their energies and interaction vertices are favorable. We focus here on $\nu_{\tau}$ CC interactions~\cite{TausInteractionsInIceCube} that have the ``Double Pulse'' and ``Double Cascade'' topologies, where a $\nu_{\tau}$ interacts with nucleon to produce a tau lepton, which then travels some distance in ice ($\propto E_\tau$) and decays into an electron or multiple hadrons.  This yields a topology with two causally-connected cascades that can give double-humped waveforms in one or more DOMs  and/or characteristic photon arrival times across multiple DOMs and strings, respectively.

This proceeding summarizes two separate analyses that exploit these aforementioned $\nu_{\tau}$ CC interaction topologies to search for $\nu_{\tau}$ and do a flavor component measurement of diffuse astrophysical neutrinos. Later, we also discuss the sensitivities of combining different event selections of IceCube to obtain the best astrophysical flavor component measurement to date. 

\section{Search for $\nu_{\tau}$-induced Double Cascades}\label{Neha}

Within 3 years of its livetime, IceCube reported evidence of an astrophysical component of diffuse neutrino flux~\cite{discovery_paper}. The event sample used for this discovery, the High Energy Starting Events (HESE), was updated over time with more years of data and updated detector and glacial ice treatment~\cite{HESE7}. This updated sample was then used to detect, for the first time, two tau neutrino candidates~\cite{Juliana} to perform a flavor composition measurement. 

Here we use the same event selection and an updated reconstruction ~\cite{Tianlu_Tables} method (compared to the one originally developed in \cite{Marcel_thesis} and further used in HESE-7.5 $\nu_{\tau}$ search \cite{Juliana}), for 12 years of IceCube HESE data, to search for ``Double Cascade'' events. The reason for using this same selection is that it is an all-sky, all-flavor selection, which also uses a self-veto technique to reject atmospheric neutrinos from the sample, with all events having reconstructed deposited energy of $E \ge 60$~TeV, giving it high astrophysical purity. We use a likelihood-based reconstruction method that compares different source hypotheses (single cascade, double cascade, and track) and assigns a topology to each event that is present in the HESE sample. PDFs of various reconstructed variables are used for flavor tagging. 
Table~\ref{analysis_variables} contains all the reconstructed variables of interest for each topology and their binning ranges and space.  In the HESE-7.5 tau search, dedicated re-simulations of two observed $\nu_{\tau}$ candidates were produced to estimate the error on the tau decay length and to estimate the probability of how compatible each event is with the background. PDFs obtained from these re-simulations were used to extend the likelihood used in ~\cite{HESE7} to improve the flavor composition measurement.  For this analysis, the variables used to generate PDFs are as indicated  in Table~\ref{analysis_variables}. A new variable, 'energy asymmetry', is included as it is a good estimator to separate single cascades from double cascades at lower reconstructed lengths. 
\begin{table}[!h]
\begin{center}

\begin{tabular}{||c | c| c| c| c||} 
 \hline
  Topology & Variable & Bin range & Number of Bins & Bin Space  \\ 
  [0.5ex] 
 \hline\hline
 \makecell {Single Cascades} & \makecell{ Energy \\  Zenith} & \makecell{60TeV - 10PeV \\ -1 - 1} & \makecell{21 \\ 10} & \makecell{$\log_{10}$ \\ cosine} \\ 
 \hline
 \makecell {Tracks} & \makecell{Energy \\  Zenith} & \makecell{60TeV - 10PeV \\ -1 - 1} & \makecell{21 \\ 10}& \makecell{$\log_{10}$ \\ cosine} \\ 
 \hline
\makecell {Double Cascades} & \makecell{ Energy \\  Length \\ Energy Asymmetry} & \makecell{60TeV - 10PeV \\ 1m - 1000m \\ -1 - 0.3 } &\makecell{21 \\ 20 \\ 26} & \makecell{$\log_{10}$ \\ $\log_{10}$ \\ linear} \\ 
 \hline
 
\end{tabular}
\caption{\label{analysis_variables} Reconstructed observables of each topology and their binning information. Reconstructed energy and zenith are the total deposited energy of and zenith of an event. Reconstructed length is the distance between vertices of two Cascades (in other words distance that tau lepton travels before it decays). Reconstructed Energy asymmetry is a measure of how the relative amount of deposited energy in each cascade is distributed and is defined as $E_A$ = $\frac{E_1 - E_2}{E_1 + E_2}$, where $E_1$ and $E_2$ are reconstructed energies of first and second cascades.}
\end{center}
\end{table}
To measure flavor fractions, a forward-folding likelihood fit is performed where, in addition to astrophysical spectral parameters such as spectral index $\gamma_{\text{astro}}$ and all flavor norm $\phi_{\text{astro}}$ (See Equation~\ref{eqn}), individual flavor fractions $f_{\alpha}$ contributing to $\phi_{\text{astro}}$ are also fitted (where $f_{\alpha}$ is fraction of $\nu_{\alpha}$ observed on earth). 
To account for atmospheric neutrino spectra, $\phi_{\text{conv}}$ (normalization of neutrino flux for the conventional component) and $\phi_{\text{prompt}}$ (normalization of neutrino flux for the prompt component)  are also considered in the fit. Detector systematics are taken into account by using the SnowStorm method ~\cite{SnowStorm_paper}, where each systematic is varied for every event while generating the Monte-Carlo simulation, allowing us to be able to combine different analysis samples in one fit. The DOM efficiency for detecting photons, the bulk ice properties such as scattering, absorption, and anisotropy, and hole ice properties (forward acceptance of DOMs due to refreezing of glacial ice after deployment of optical modules) are used as nuisance parameters in the fit.

\subsection{HESE-12 Flavor Fit}\label{HESEonlyFit}
In this section, we discuss the Asimov sensitivity to constrain flavor contours using 12 years of IceCube HESE data. The astrophysical spectrum is modeled here as a single power-law with flavor ratio $\nu_{e}:\nu_{\mu}:\nu_{\tau} = 1:1:1$,  following a measurement from a sample taken with 9.5 years of IceCube data ~\cite{10diffuse},
\begin{equation} \label{eqn}
\phi(E) = \phi_{\text{astro}}\bigg(\frac{E}{\text{100 TeV}}\bigg)^{-\gamma_{\text{astro}}}10^{-18}{\text{ GeV}}^{-1}{\text{cm}}^{-2}{\text{s}}^{-1}{\text{sr}}^{-1}
\end{equation}
with a total normalization of $\phi_{\text{astro}}$ = 4.32 and spectral index $\gamma_{\text{astro}}$ = 2.37.
\begin{figure}[ht]
  \centering
    \includegraphics[width=0.6\columnwidth]{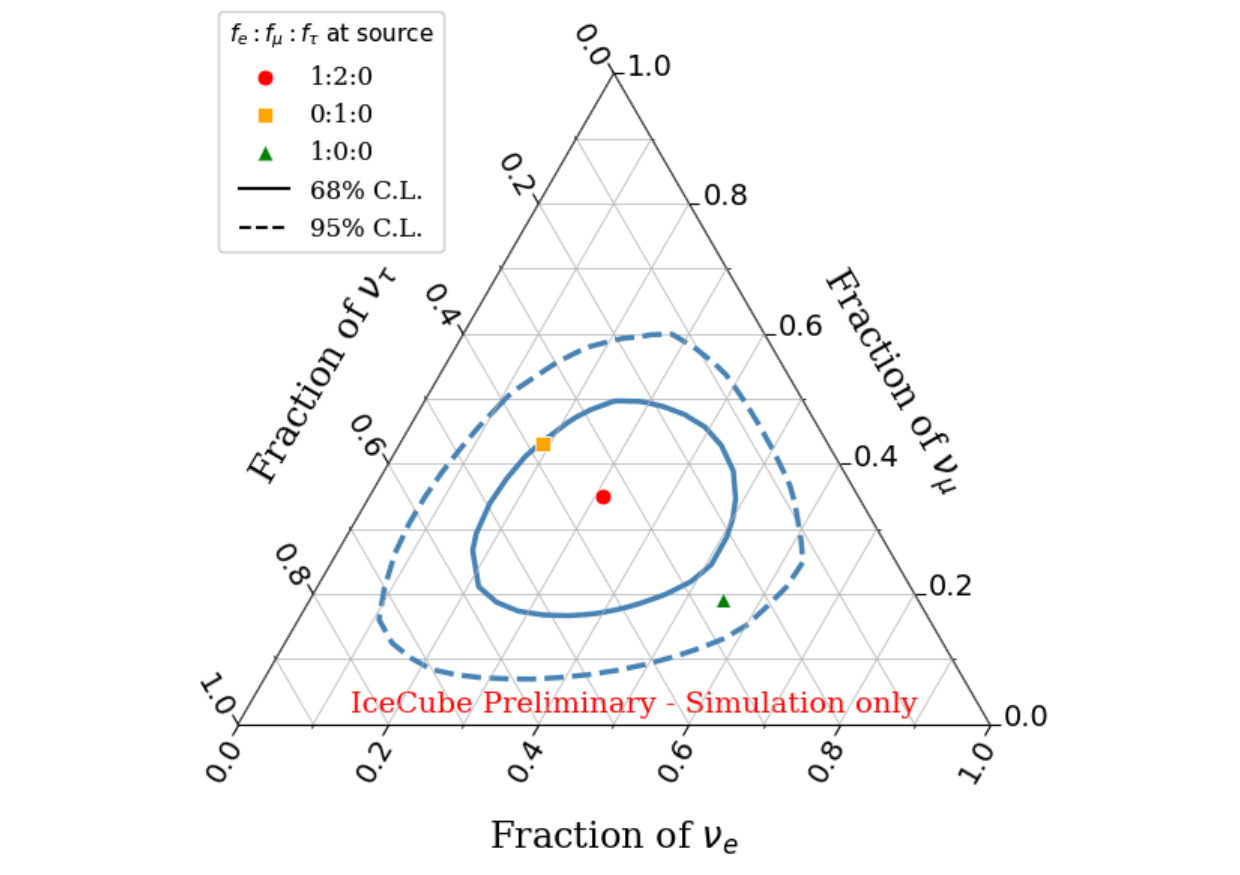}
    \caption{Asimov sensitivity of this analysis to measure flavor composition using 12 years of IceCube HESE data, for an astrophysical neutrino spectrum following the single unbroken power-law given in Equation \ref{eqn}, with a flavor composition of $\nu_{e}:\nu_{\mu}:\nu_{\tau} = 1:1:1$.}
    \label{fig:HESE-12}

\vspace{6mm}

  \centering
    \includegraphics[width=0.6\columnwidth]{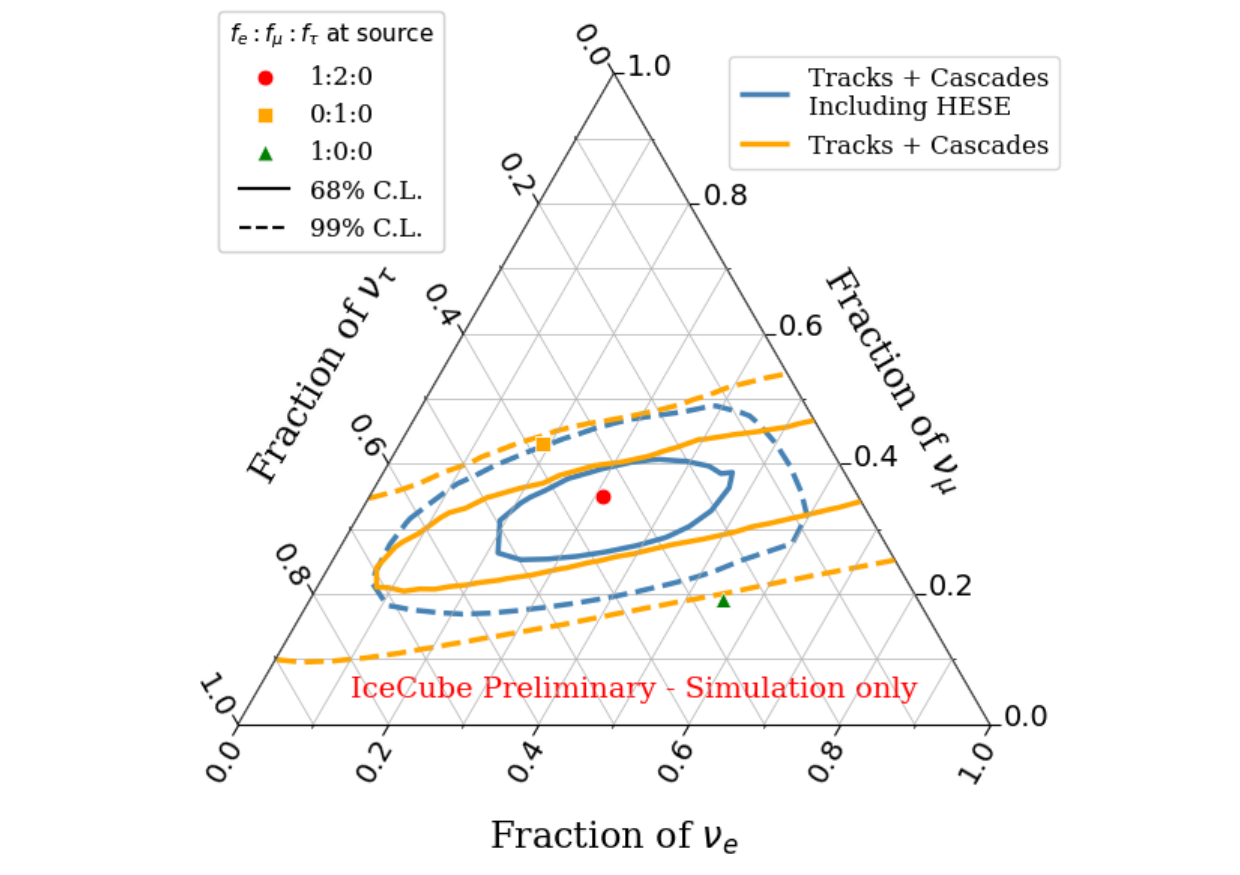}
    \caption{Asimov sensitivity of this analysis to measure flavor composition by combining 12 years of HESE sample with current GlobalFit sample~\cite{GlobalFit:2023icrc}, for an astrophysical neutrino spectrum following the single unbroken power-law given in Equation \ref{eqn}, with a flavor composition of $\nu_{e}:\nu_{\mu}:\nu_{\tau} = 1:1:1$.}
    \label{fig:Combined}
\end{figure}

\subsection{Towards a GlobalFit of flavor Measurement}\label{GlobalFit}

The ultimate goal of this analysis is to measure the flavor content of diffuse astrophysical neutrinos using not just the HESE sample but doing so by combining it with other event samples. The HESE sample, as pure as it may be, does not contain enough events to tightly constrain the $\nu_{\mu}$ and $\nu_{e}$ fractions. The idea of a GlobalFit is to combine samples that individually have their advantages and disadvantages in a combined fit to model all nuisance and signal parameters in a self-consistent way. In the current GlobalFit scheme of IceCube, it is shown that this cogent treatment of all parameters in a fit is not only possible but also gives the most precise measurement of the astrophysical neutrino spectrum (assuming single power law)  made to date ~\cite{GlobalFit:2023icrc}. The addition of a $\nu_{\tau}$ identifier is capable of excluding the neutron beam scenario ($\nu_{e}:\nu_{\mu}:\nu_{\tau} = 1:0:0$ at the source) with >3$\sigma$ confidence if the injected source mechanism is the Pion-Production scenario ($\nu_{e}:\nu_{\mu}:\nu_{\tau} = 1:2:0$ at the source).

As can be seen from Fig.\ref{fig:HESE-12} and Fig.~\ref{fig:Combined}, it is clear that the HESE-only fit is sensitive to constraining the $\nu_{\tau}$ fraction very well but fails to do so for the $\nu_{\mu}$ fraction. This is understandable as in the HESE sample, the number of distinguishable $\nu_{\mu}$ events is low. This, in turn, can be provided by Tracks from GlobalFit.

\section{Observation of Seven Astrophysical $\nu_{\tau}$ Candidates}

\begin{figure}[ht]
\begin{minipage}{.48\textwidth}
    \centering
    \includegraphics[width=1.0\columnwidth]{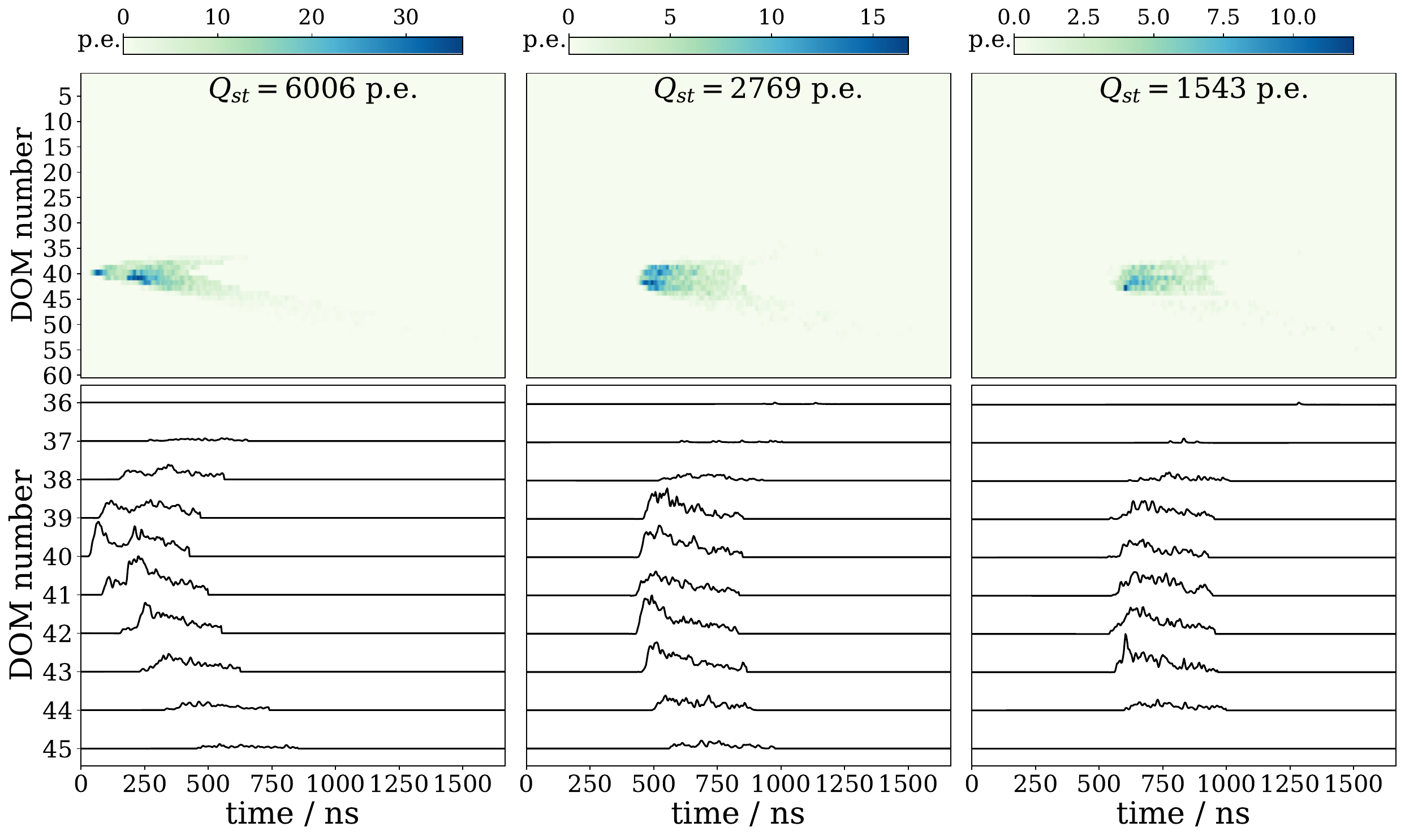}
    \caption{Candidate astrophysical $\nu_{\tau}$ detected in Nov. 2019.  Each column corresponds to a string in the selected event.  The top row of figures shows the DOM number (proportional to depth) versus the time of the digitized PMT signal in 3~ns bins, with the color scale giving the signal amplitude in p.e.~in each time bin. The total p.e. detected on each string, $Q_\mathrm{str.}$, is shown.  The bottom row shows the digitized waveforms for a subset of the DOMs on each string.  The amplitudes of the waveforms in each string are normalized to the peak amplitude in that string.}
\label{fig:NuTauCandidate}
\end{minipage} \hfill
\begin{minipage}{.48\textwidth}
  \centering
    \includegraphics[width=1.0\columnwidth]{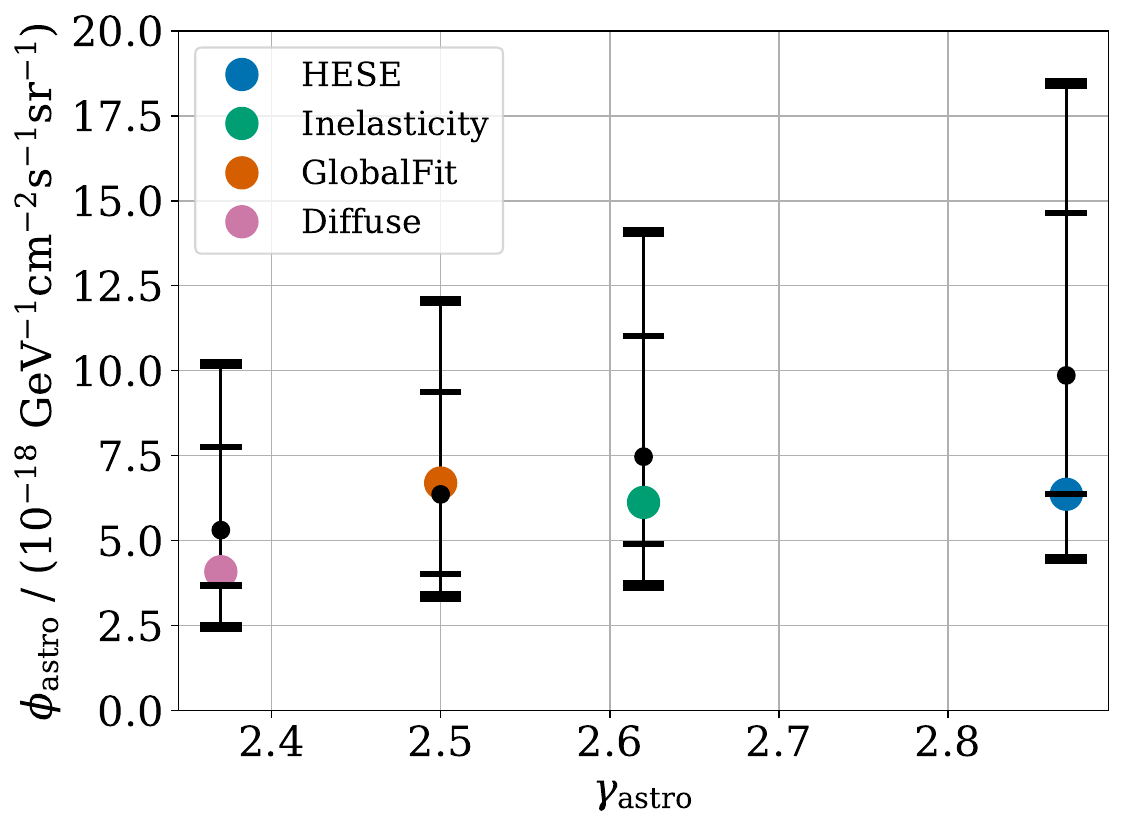}
    \caption{Measured $\nu_{\tau}^\mathrm{astro}$ flux normalizations (black dots) with 68\% and 90\% confidence intervals (black error bars), for each of $\phi_\mathrm{astro}^\mathrm{IC}$, denoted by colored circles and labeled as ``HESE''~\cite{IceCube:2020wum}, ``Inelasticity''~\cite{IceCube:2018pgc}, ``GlobalFit''~\cite{IceCube:2015gsk} and ``Diffuse''~\cite{IceCube:2021uhz}.  The spectral index $\gamma_\mathrm{astro}$ was not measured by this analysis. The fluxes shown have each been normalized to their all-flavor values under the assumption of a 1:1:1 flavor ratio at Earth.}
\label{fig:flux_fits}
\end{minipage}
\end{figure}

In a second approach, we used convolutional neural networks (CNNs) that had been trained on images derived from both simulated $\nu_\tau$ events and simulated background events.  These images, shown for one of the candidate events in the top row of Fig.~\ref{fig:NuTauCandidate}, encapsulated the full waveform information from up to 180 DOMs on three neighboring strings centered on the string with the highest charge in the event.  Based on simulations, a median detected neutrino energy of approximately 200~TeV is predicted for $\nu_\tau$ events, which assume an astrophysical neutrino flux from Ref.~\cite{IceCube:2020wum}. These simulations indicate that the events exhibit energies ranging from roughly 20~TeV to 1~PeV.  Combined with a few other selection criteria, the CNNs identified seven candidate $\nu_\tau$ events.  Considering backgrounds from astrophysical neutrinos, conventional atmospheric neutrinos, conventional atmospheric muons, and prompt atmospheric neutrinos, we obtain a total estimated background of roughly 0.5 events, depending on the assumed astrophysical neutrino flux.  The dominant background contribution is from non-$\nu_\tau$ astrophysical neutrinos, and we assumed the prompt muon flux was zero.  In the context of these backgrounds, we are thus able to exclude the absence of astrophysical $\nu_\tau$ at the $5\sigma$ level. We also measure the astrophysical $\nu_\tau$ flux, finding that it is consistent with expectations based on previously published IceCube astrophysical neutrino flux measurements~\cite{IceCube:2020wum,IceCube:2018pgc,IceCube:2015gsk,IceCube:2021uhz}.

One of the seven candidate $\nu_\tau$ events is shown in Fig.~\ref{fig:NuTauCandidate}.  While this event shows some evidence of DOM waveforms with the double pulse signature, studies where individual waveforms were smoothed showed that the CNN scores were sensitive to the overall event structure rather than local double-pulse waveforms of individual DOMs.  Figure~\ref{fig:flux_fits} shows the best-fit tau neutrino fluxes (black dots) for four IceCube-measured astrophysical fluxes (colored dots).  The inner (outer) bars denote the frequentist 68\% (90\%) confidence intervals.  The best-fit points from the respective IceCube analyses are each within the 68\% confidence intervals of this analysis.

\begin{figure}[h]
\centering
\end{figure}

\section{Conclusions and Outlook}\label{conclusions}
A dedicated study to look for tau neutrinos is needed to do a flavor measurement of astrophysical neutrinos that reach Earth. The fraction of each flavor is related to the population of their sources and their neutrino production mechanisms. When a tau neutrino has a CC interaction in IceCube, a tau lepton is created, which upon traveling some distance, often decays to create another shower of particles, giving two causally connected light depositions, producing a double cascade and/or double pulse signature.

Here we described two methods in IceCube used to identify these $\nu_{\tau}$ events, building on previous IceCube searches~\cite{PhysRevD.86.022005,PhysRevD.93.022001,Meier:2019ypu}. The first approach was used in the 7.5-year HESE sample~\cite{IceCube:2020fpi} to find two double-cascade $\nu_{\tau}$ candidates and performed a flavor composition measurement that yielded the first-ever non-zero $\nu_{\tau}$ fraction measured at earth. In this proceeding, we discussed the IceCube sensitivity achieved by extending this sample to 12 years, together with a novel treatment of detector systematics in the fit, to deliver tighter constraints on the flavor measurement. We also showed that if this tau identifier is combined with other high statistics samples of cascades and tracks, the analysis becomes capable of rejecting the neutron beam neutrino production mechanism at greater than 3$\sigma$.
The second approach using CNNs made the most highly-significant rejection to date of the null hypothesis of zero astrophysical tau neutrinos.

\bibliographystyle{ICRC}
\bibliography{references}

%

\clearpage

\section*{Full Author List: IceCube Collaboration}

\scriptsize
\noindent
R. Abbasi$^{17}$,
M. Ackermann$^{63}$,
J. Adams$^{18}$,
S. K. Agarwalla$^{40,\: 64}$,
J. A. Aguilar$^{12}$,
M. Ahlers$^{22}$,
J.M. Alameddine$^{23}$,
N. M. Amin$^{44}$,
K. Andeen$^{42}$,
G. Anton$^{26}$
% ...
\\
\\
% ...
$^{12}$ Universit{\'e} Libre de Bruxelles, Science Faculty CP230, B-1050 Brussels, Belgium \\
% ...
$^{17}$ Department of Physics, Loyola University Chicago, Chicago, IL 60660, USA \\
$^{18}$ Dept. of Physics and Astronomy, University of Canterbury, Private Bag 4800, Christchurch, New Zealand \\
% ...
$^{22}$ Niels Bohr Institute, University of Copenhagen, DK-2100 Copenhagen, Denmark \\
$^{23}$ Dept. of Physics, TU Dortmund University, D-44221 Dortmund, Germany \\
% ...
$^{26}$ Erlangen Centre for Astroparticle Physics, Friedrich-Alexander-Universit{\"a}t Erlangen-N{\"u}rnberg, D-91058 Erlangen, Germany \\
% ...
$^{40}$ Dept. of Physics and Wisconsin IceCube Particle Astrophysics Center, University of Wisconsin{\textendash}Madison, Madison, WI 53706, USA \\
% ...
$^{42}$ Department of Physics, Marquette University, Milwaukee, WI, 53201, USA \\
% ...
$^{44}$ Bartol Research Institute and Dept. of Physics and Astronomy, University of Delaware, Newark, DE 19716, USA \\
% ...
$^{63}$ Deutsches Elektronen-Synchrotron DESY, Platanenallee 6, 15738 Zeuthen, Germany  \\
$^{64}$ Institute of Physics, Sachivalaya Marg, Sainik School Post, Bhubaneswar 751005, India \\
% ...

\subsection*{Acknowledgements}

\noindent
The authors gratefully acknowledge the support from the following agencies and institutions:
USA {\textendash} U.S. National Science Foundation-Office of Polar Programs,
U.S. National Science Foundation-Physics Division,
U.S. National Science Foundation-EPSCoR,
Wisconsin Alumni Research Foundation,
Center for High Throughput Computing (CHTC) at the University of Wisconsin{\textendash}Madison,
Open Science Grid (OSG),
Advanced Cyberinfrastructure Coordination Ecosystem: Services {\&} Support (ACCESS),
Frontera computing project at the Texas Advanced Computing Center,
U.S. Department of Energy-National Energy Research Scientific Computing Center,
Particle astrophysics research computing center at the University of Maryland,
Institute for Cyber-Enabled Research at Michigan State University,
and Astroparticle physics computational facility at Marquette University;
Belgium {\textendash} Funds for Scientific Research (FRS-FNRS and FWO),
FWO Odysseus and Big Science programmes,
and Belgian Federal Science Policy Office (Belspo);
Germany {\textendash} Bundesministerium f{\"u}r Bildung und Forschung (BMBF),
Deutsche Forschungsgemeinschaft (DFG),
Helmholtz Alliance for Astroparticle Physics (HAP),
Initiative and Networking Fund of the Helmholtz Association,
Deutsches Elektronen Synchrotron (DESY),
and High Performance Computing cluster of the RWTH Aachen;
Sweden {\textendash} Swedish Research Council,
Swedish Polar Research Secretariat,
Swedish National Infrastructure for Computing (SNIC),
and Knut and Alice Wallenberg Foundation;
European Union {\textendash} EGI Advanced Computing for research;
Australia {\textendash} Australian Research Council;
Canada {\textendash} Natural Sciences and Engineering Research Council of Canada,
Calcul Qu{\'e}bec, Compute Ontario, Canada Foundation for Innovation, WestGrid, and Compute Canada;
Denmark {\textendash} Villum Fonden, Carlsberg Foundation, and European Commission;
New Zealand {\textendash} Marsden Fund;
Japan {\textendash} Japan Society for Promotion of Science (JSPS)
and Institute for Global Prominent Research (IGPR) of Chiba University;
Korea {\textendash} National Research Foundation of Korea (NRF);
Switzerland {\textendash} Swiss National Science Foundation (SNSF);
United Kingdom {\textendash} Department of Physics, University of Oxford.

\end{document}